\documentclass[runningheads]{llncs}
\usepackage[T1]{fontenc}
\usepackage{graphicx}
\usepackage{caption}
\usepackage{subcaption}
\usepackage{amsmath}
\usepackage{booktabs}
\usepackage{siunitx}
\usepackage{bbm}
\PassOptionsToPackage{hyphens}{url}\usepackage{hyperref}

\begin{document}
\title{4D Pencil Beam Treatment Plan with Conditional Weight Predictions}
%
%\titlerunning{Abbreviated paper title}
% If the paper title is too long for the running head, you can set
% an abbreviated paper title here
%

\author{Nair N von Mühlenen\inst{1}\orcidID{0009-0002-5122-5480} \and Florentin Bieder\inst{1}\orcidID{0000-0001-9558-0623} \and Philippe C Cattin\inst{1}\orcidID{0000-0001-8785-2713}}

\authorrunning{N. von Mühlenen et al.}
%First names are abbreviated in the running head.
% If there are more than two authors, 'et al.' is used.
\institute{ Department of Biomedical Engineering, University of Basel, 4001 Basel, Switzerland \\
\email{nair.vonmuehlenen@unibas.ch}\\}

%\author{Anonymized Authors}
%\authorrunning{Anonymized Author et a}
%\institute{Anonymized Affiliations\\
%\email{email@anonymized.com}\\}
%
\maketitle              % typeset the header of the contribution
\begin{abstract}
\textit{Objective}. This work aims to accelerate our four-dimensional (4D) proton pencil beam delivery strategy, which incorporates respiratory motion into a dynamic treatment plan, to improve dose conformity and treatment efficiency for gantry-less and magnet-free scanner designs.
\textit{Approach}. To accelerate the generation and adaptation of 4D treatment plans, we propose a conditional hybrid ResNet18-Transformer model for predicting pencil beam weights. The model predicts either the full set of pencil beam weights or a subset thereof.
\textit{Main Results}. The conditional model can successfully predict the remaining beam weights of a treatment, and the results suggest that extending this approach to full treatment prediction is feasible but requires further investigation. 
\textit{Significance}. The acceleration of 4D treatment generation via pencil beam weight prediction takes us one step closer to the feasibility of treating mobile targets with simplified, gantry-less, and magnet-free scanner designs. Reducing system complexity while preserving dosimetric conformity may offer a pathway toward more accessible and cost-effective proton beam therapy for motion-affected tumours. Our code is available on \href{https://github.com/NairVonMuehlenen/Pencil-Beam-Weight-Prediction}{github.com/NairVonMuehlenen/Pencil-Beam-Weight-Prediction}.

\end{abstract}

\section{Introduction}
Proton therapy (PT) is an advanced form of radiation therapy used in the treatment of cancer \cite{smith2006proton}. Unlike conventional photon-based radiotherapy, PT uses protons to deliver radiation. An advantage of this treatment stems from the unique physical properties of protons, particularly the Bragg Peak, which enable highly localised radiation deposition \cite{schulz2007particle}. Beyond this peak, the radiation dose drops rapidly, so virtually no radiation is delivered beyond the target depth. This characteristic dose distribution provides significant advantages for cancer treatment as it can minimise radiation exposure to surrounding healthy tissue, organs at risk (OAR) and other sensitive structures \cite{mohan2022review,mohan2017proton}. Despite this significant advantage, PT also presents several challenges. The treatment is highly sensitive to patient and tumour motion, which can affect the placement of the delivered dose. In addition, the infrastructure required for PT is complex and expensive, limiting its availability to specialised centres and making it inaccessible to many worldwide. 

\begin{figure}[h]
    \centering
    \includegraphics[width=1\linewidth]{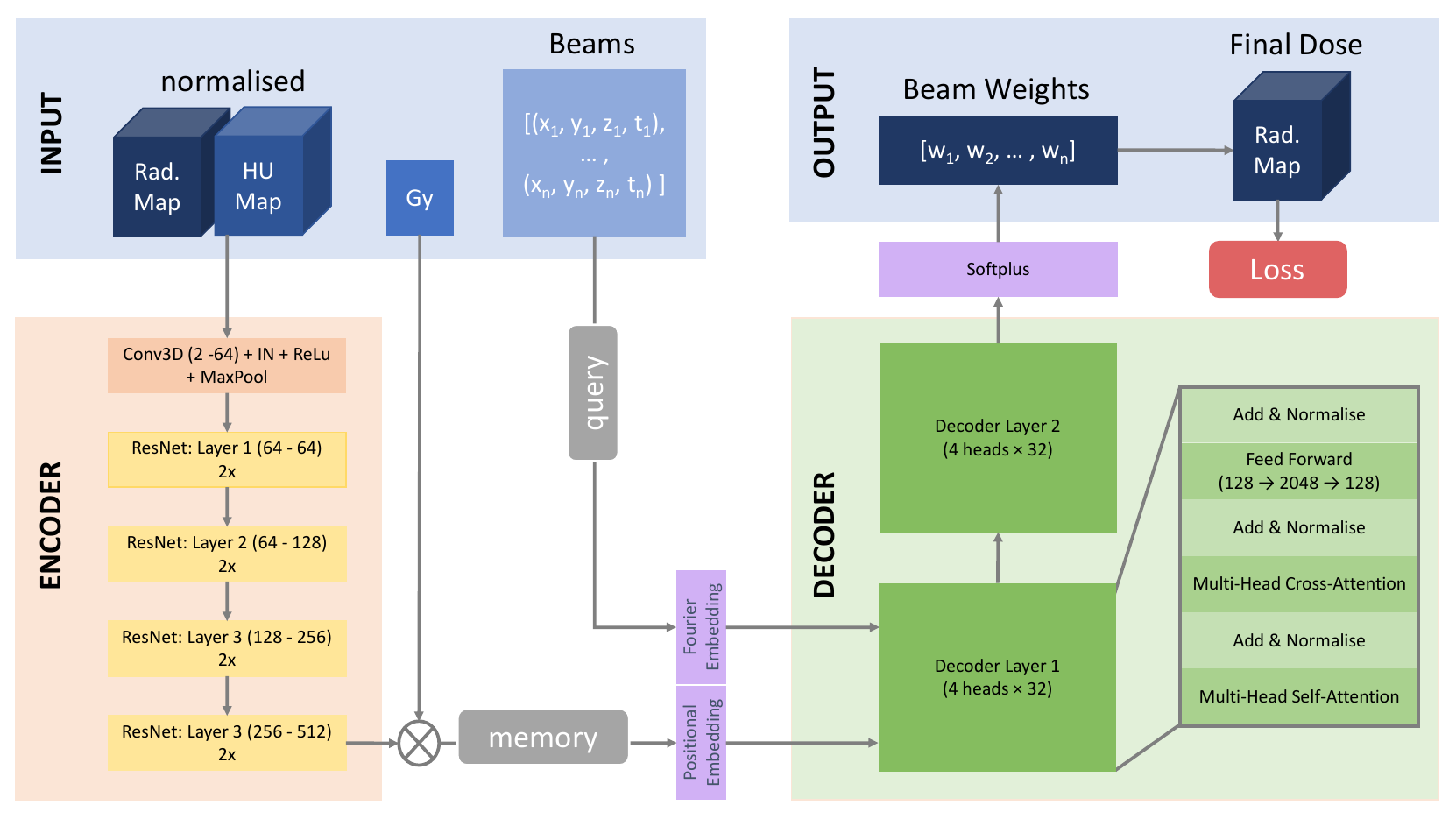}
    \caption{Schematic diagram of the conditional hybrid ResNet18–Transformer: It consists of a ResNet18 encoder and a Transformer decoder. The inputs to the model are the motion-deformed CT scan with the current accumulated dose distribution, the prescribed Gray (PG) and a predetermined sequence of spots. }\label{fig:modell}
\end{figure}

This work of von Mühlenen et al. \cite{vonmuehlenen2026magnetfreeprotontherapy4d} introduced a new 4D planning framework for mobile tumours and its possible application in gantry-less and even magnet-less PT systems. It showed promising future applications of more cost-effective gantry-less PT systems. The system takes as input a CT scan, the planning target volume (PTV), and the corresponding respiratory motion fields, and outputs the pencil beam positions, their delivery times relative to the start of treatment, and their weights. The delivery time is composed of a nominal dwell time and the travel time between spots. The beam weights are optimised to achieve homogeneous dose coverage of the target while minimising radiation exposure to healthy tissue. The optimisation minimises an objective function composed of common PT metrics such as $\mathrm{V_{95 \%}}$ and $\mathrm{D_{95 \%}}$. With the optimised pencil beam weights, the treatment plan can be adapted. In particular, removing spots with negligible weights and adapting the nominal dwell time to the actual dwell time, altering the temporal structure of the delivery sequence.
These properties motivate an iterative planning strategy, in which beam weights are recomputed to refine and potentially simplify the delivery plan. However, the conventional weight optimisation is too slow for such an application. This motivated the development of a fast weight-prediction model that enables rapid recomputation and adjustment of treatment plans while maintaining treatment-plan quality.  An overview of the proposed architecture is shown in Fig.~\ref{fig:modell}.

\subsection{Related Work}
The recent years have also seen significant progress in 4D treatment planning for PT, especially for moving targets such as lung tumours. Modern approaches now enable robust, patient-specific dose calculations that account for respiratory motion and anatomical changes, improving treatment accuracy and safety \cite{Liu2016Exploratory,Meijers2019Log}. Several centres have implemented clinically feasible, custom 4D dose reconstruction and accumulation tools within commercial planning systems, allowing for retrospective and prospective plan evaluation and supporting adaptive plan evaluation with processing times of approximately 10 minutes per treatment fraction and only limited manual intervention \cite{Lebbink2023Parameter,Meijers2019Log}. Four-dimensional robust optimisation (4DRO) strategies have demonstrated improved target coverage and reduced interplay effects compared to traditional 3D planning \cite{Liu2016Exploratory,Mastella20204D}. Adaptive planning methods, such as dose-mimicking and template-based approaches, are being explored to automate plan adaptation during treatment, thereby further enhancing robustness and sparing OAR \cite{Kaushik2024Adaptive,Lebbink2023Parameter}. 

Despite these advances, several challenges remain.
(1) Computational Demands: Full 4D robust optimisation remains computationally intensive, though strategies to reduce the number of phases or use surrogate phases have been proposed to balance speed and quality \cite{Kaushik2024Adaptive,Mastella20204D}. 
(2) Motion and Anatomy Variability: Accurate modelling of patient-specific motion and anatomical changes over the treatment course is complex and requires repeated imaging or advanced motion modelling \cite{Lebbink2023Parameter,Meijers2019Log}. 
(3) Clinical Integration: While proof-of-concept and early clinical implementations exist, large-scale, fully automated, and prospective 4D adaptive workflows are not yet standard in most clinics \cite{Kaushik2024Adaptive,Meijers2019Log}.

\section{Method}
Since weight optimisation is the most computationally expensive step in the 4D treatment planning process, we propose a significantly faster weight-prediction model for treatment adaptation and, potentially, for full treatment beam weight prediction. Specifically, we employ a conditional hybrid ResNet18-Transformer model that predicts either the full set of pencil beam weights or a subset thereof. To ensure comparable plan quality, the model is trained using the same metric-based loss function as the optimisation framework shown in von Mühlenen et al. \cite{vonmuehlenen2026magnetfreeprotontherapy4d}, promoting homogeneous target coverage while minimising dose to healthy tissue.

\subsection{Dataset}
\begin{figure}[h]
    \centering
    \includegraphics[width=1\linewidth]{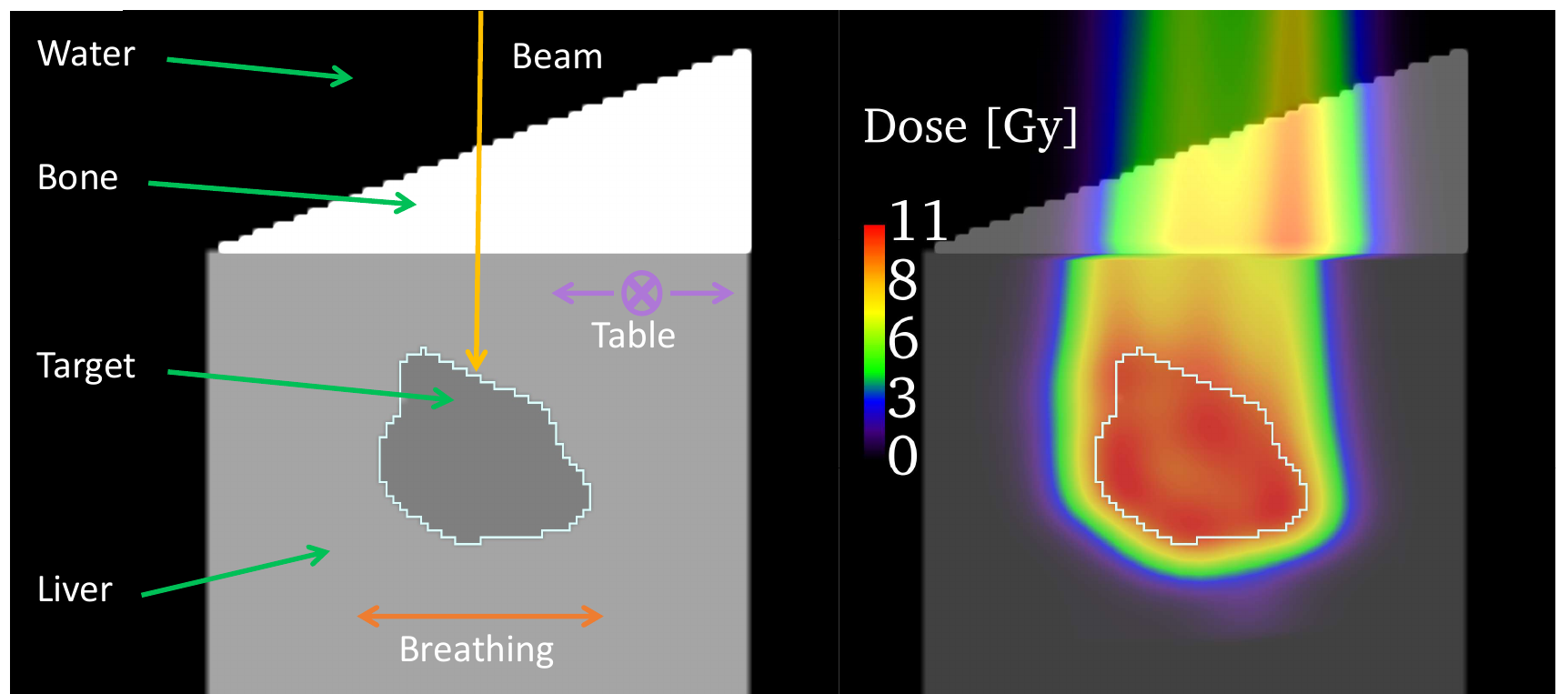}
    \caption{Example of a simulated phantom: On the right, a CT centre slice showing heterogeneous tissue-equivalent regions. On the left, the corresponding radiation map of the centre slice with optimised weights. }\label{fig:HU}
\end{figure}

In this study, we use simulated data representing a liver tumour under motion. This choice provides full control over the entire workflow and ensures that all components can be evaluated under controlled, reproducible conditions. The data generation pipeline follows the aforementioned work \cite{vonmuehlenen2026magnetfreeprotontherapy4d}, except that we extend the dataset from spherical to more anatomically realistic target shapes. For these additional geometries, corresponding optimised beam weights are available. In the simulated data, the phantom patient is composed of different tissue-equivalent regions with varying densities and thicknesses. Figure~\ref{fig:HU} shows a CT slice from simulated data on the right and the corresponding dose distribution with optimised beam weights on the left. For each phantom, we compute a set of planned pencil beam positions and possible delivery times.

\subsection{Model Architecture}
The proposed conditional hybrid ResNet18–Transformer model consists of a ResNet18 encoder \cite{he2016deep} and a Transformer decoder \cite{vaswani2017attention}. The encoder receives the motion-deformed CT scan along with the current accumulated dose distribution, which represents the dose delivered up to the current point in time. The extracted feature representation is concatenated with the prescribed dose (PG) and provided to the Transformer decoder. The decoder receives, as queries, a predetermined sequence of Fourier-embedded beam positions and delivery times, ${ (x_i, y_i, z_i, t_i) }_{i=1}^N$. Depending on the application, the query sequence may include either the full set of planned spots or only the remaining delivery sequence from a given time onward. In the latter case, the accumulated dose supplied to the encoder corresponds to the dose delivered up to the first queried beam. The decoder predicts one weight per queried pencil beam. A schematic overview of the proposed architecture is shown in Fig.~\ref{fig:modell}.
\newpage
\subsection{Training Objective}
Rather than directly regressing the optimised beam weights, the proposed network is trained to minimise the objective function used during treatment plan optimisation. From the predicted beam weights, we calculate the dose distribution within the phantom using the analytical model of Schaffner et al. \cite{schaffner1999dose}. The resulting distribution is then evaluated using the same objective function employed in the optimisation framework von Mühlenen et al. \cite{vonmuehlenen2026magnetfreeprotontherapy4d}. This ensures that training directly optimises treatment plan quality instead of agreement with a particular set of beam weights. The loss is formulated in terms of the dosimetric metrics $\mathrm{V_{95 \%}}$, $\mathrm{D_{95 \%}}$, $\mathrm{D_{max}TV}$ and $\mathrm{D_{max}HT}$ and is defined as

\begin{equation}\label{eq:Loss}
    \mathcal{L} =\mathcal L_V + \mathcal L_D + c_{max} \left( \mathcal L_{DmaxTV} + \mathcal L_{DmaxHT} \right), 
\end{equation}
where $c_{max}$ controls the relative contribution of the maximum-dose penalties. The individual loss terms are given in Equations~\ref{eq:Loss_V} - \ref{eq:Loss_maxHT}, while the corresponding optimisation metrics are summarised in Table~\ref{tab:todo}. Each loss component is computed relative to the prescribed physical dose.

\begin{align}
    \mathcal L_V &= \max\{\mathrm{L}_V - \mathrm{V_{95\%}}, 0\} \label{eq:Loss_V}
    \\
    \mathcal L_D &= \max\{\mathrm{L}_D-\mathrm{D_{95\%}}, 0\} \label{eq:Loss_D}
    \\
    \mathcal L_{DmaxTV} &= \max\{\mathrm{D}_{max}\mathrm{TV} - \mathrm{L}_{max}\mathrm{TV}, 0\} \label{eq:Loss_maxTV}
    \\
    \mathcal L_{DmaxHT} &= \mathrm{D}_{max} \mathrm{HT} - \mathrm{L}_{max} \mathrm{HT} \label{eq:Loss_maxHT}
\end{align}

\begin{table*}[h]
\centering
\caption{Used optimisation metrics depended on the prescribed Gray (PG).}\label{tab:todo}
    \begin{tabular}{llc}
        \toprule
        \textbf{Notations} & \textbf{Definition} & \textbf{Optimisation Metric}\\
        \midrule
        $\mathrm{L}_{V}$ & The ideal value for $\text{V}_{95\%}$ &\SI{98}{\%} of PG \\ 
        $\mathrm{L}_{D}$ & The ideal value for $\text{D}_{95\%}$ & PG\\ 
        $\mathrm{L}_{max}\mathrm{TV}$ & The maximum for $\text{D}_{max}\text{TV}$ & $<$ \SI{120}{\%} of PG\\ 
        $\mathrm{L}_{max}\mathrm{HT}$ & The maximum for $\text{D}_{max}\text{HT}$ & $<$ \SI{70}{\%} of PG\\ 
        \bottomrule
    \end{tabular}
\end{table*}

\subsection{Training Procedure}
The network was trained using the Adam optimiser \cite{kingma2014adam} with a learning rate of $10^{-5}$ and a batch size of 2 for 70 epochs, resulting in approximately 17600 optimisation steps per epoch. The training was performed on a dataset comprising 20 simulated phantom patients, with one patient removed for testing and the remaining 19 split into 5 folds for training and validation. The dataset is augmented to include prescribed physical Grey doses ranging from 1 to \SI{80}{Gy}. The loss weight $c_{max}$ was set to 0.5. All experiments were conducted on a single NVIDIA A100 GPU.

\begin{figure}[tb]
    \centering
    \includegraphics[width=1\linewidth]{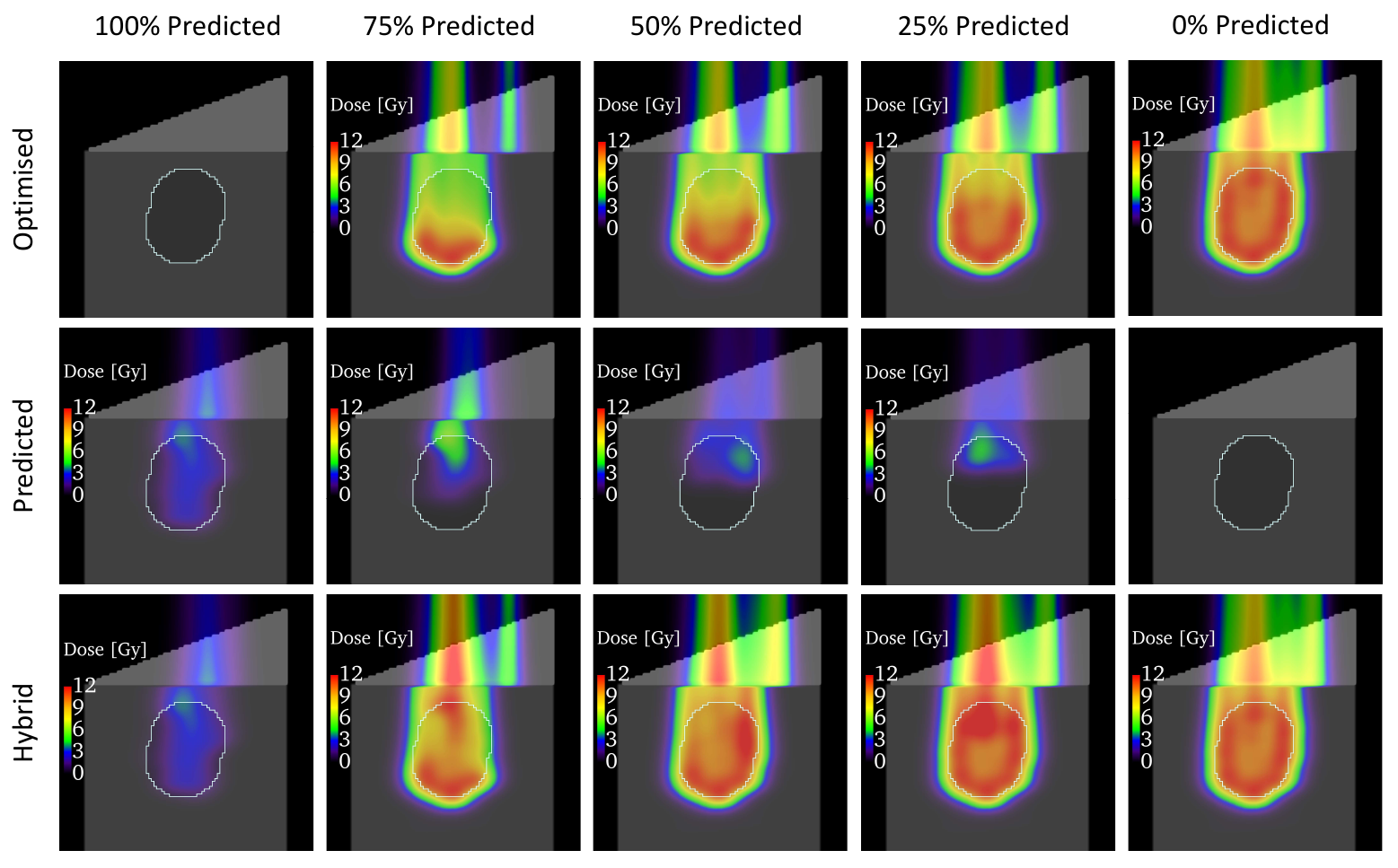}
    \caption{Dose maps for full- and partial-beam weight prediction. Columns, from left to right, correspond to the fully predicted treatment plan (100\%), the remaining \SI{75}{\%}, \SI{50}{\%}, and \SI{25}{\%} of the treatment, and the ground truth. The first row shows the accumulated partial dose distribution used as input to the model; the second row shows the dose contribution from the predicted beam weights; and the third row shows the hybrid treatment plan obtained by combining the reference with the predicted beam weights.}\label{fig:results}
\end{figure}

\section{Results}
We evaluated the proposed model by predicting both the complete set of beam weights and the remaining beam weights at different stages of treatment delivery. For the evaluation, we utilised \SI{10}{Gy} as the PG. Figure~\ref{fig:results} compares the corresponding dose distributions for the fully predicted, partially predicted, and fully optimised ground truth (GT) treatment plans. The first row illustrates the accumulated dose distribution provided as input to the model, obtained from the GT plan. The second row presents the dose distributions resulting from the predicted beam weights, while the final row shows the combined distributions obtained by merging the optimised and predicted beam weights. These hybrid dose distributions are used to evaluate model performance using the loss defined in Equation~\ref{eq:Loss}. Partial predictions were evaluated after \SI{25}{\%}, \SI{50}{\%}, and \SI{75}{\%} of the treatment had been delivered, corresponding to predictions for the remaining \SI{75}{\%}, \SI{50}{\%}, and \SI{25}{\%} of the treatment, respectively. As expected, prediction accuracy improved as the proportion of beam weights requiring prediction decreased. The closest agreement with the fully optimised reference was achieved when only the final \SI{25}{\%} of the treatment required prediction, followed by prediction of the remaining \SI{50}{\%} and \SI{75}{\%} of the treatment. Predicting the complete set of beam weights resulted in the largest deviations from the optimised reference.

\begin{figure}[htbp]
    \centering
    \begin{subfigure}{0.5\textwidth}
        \centering
        \includegraphics[width=\textwidth]{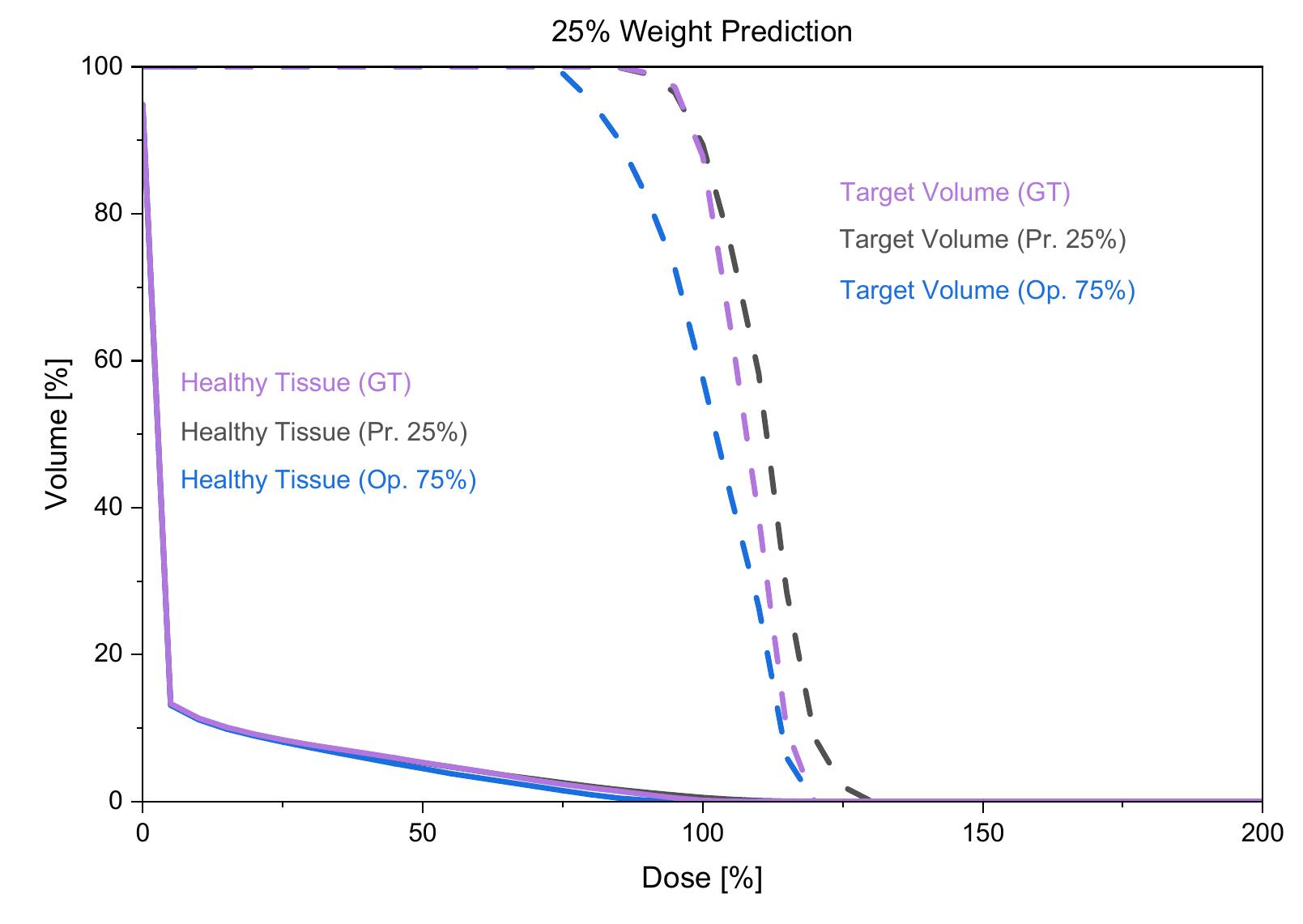}
        \caption{}\label{fig:DVH25}
    \end{subfigure}%
    \hfill 
    \begin{subfigure}{0.5\textwidth}
        \centering
        \includegraphics[width=\textwidth]{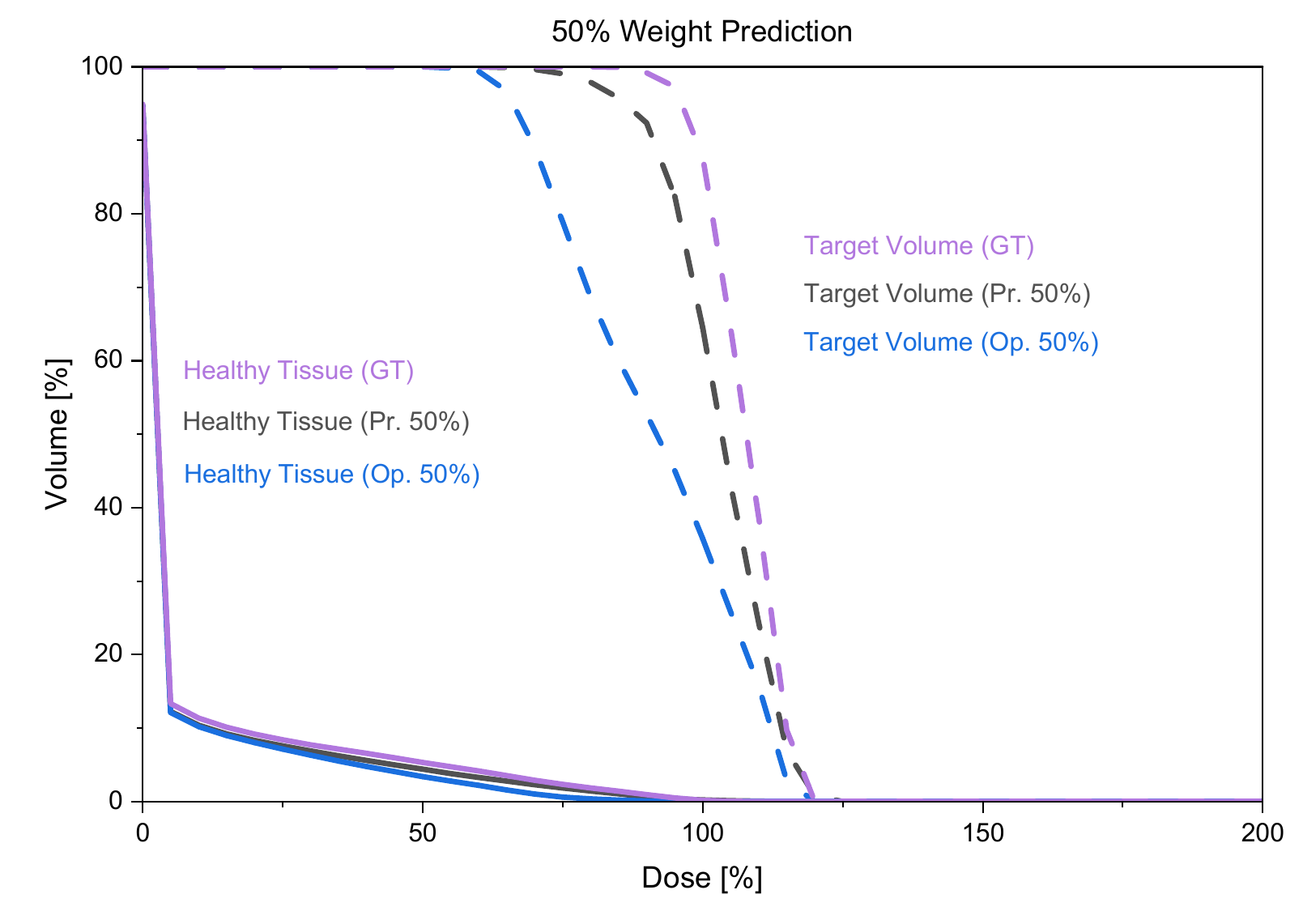}
        \caption{}\label{fig:DVH50}
    \end{subfigure}
    \vspace{1em} 
    \begin{subfigure}{0.5\textwidth}
        \centering
        \includegraphics[width=\textwidth]{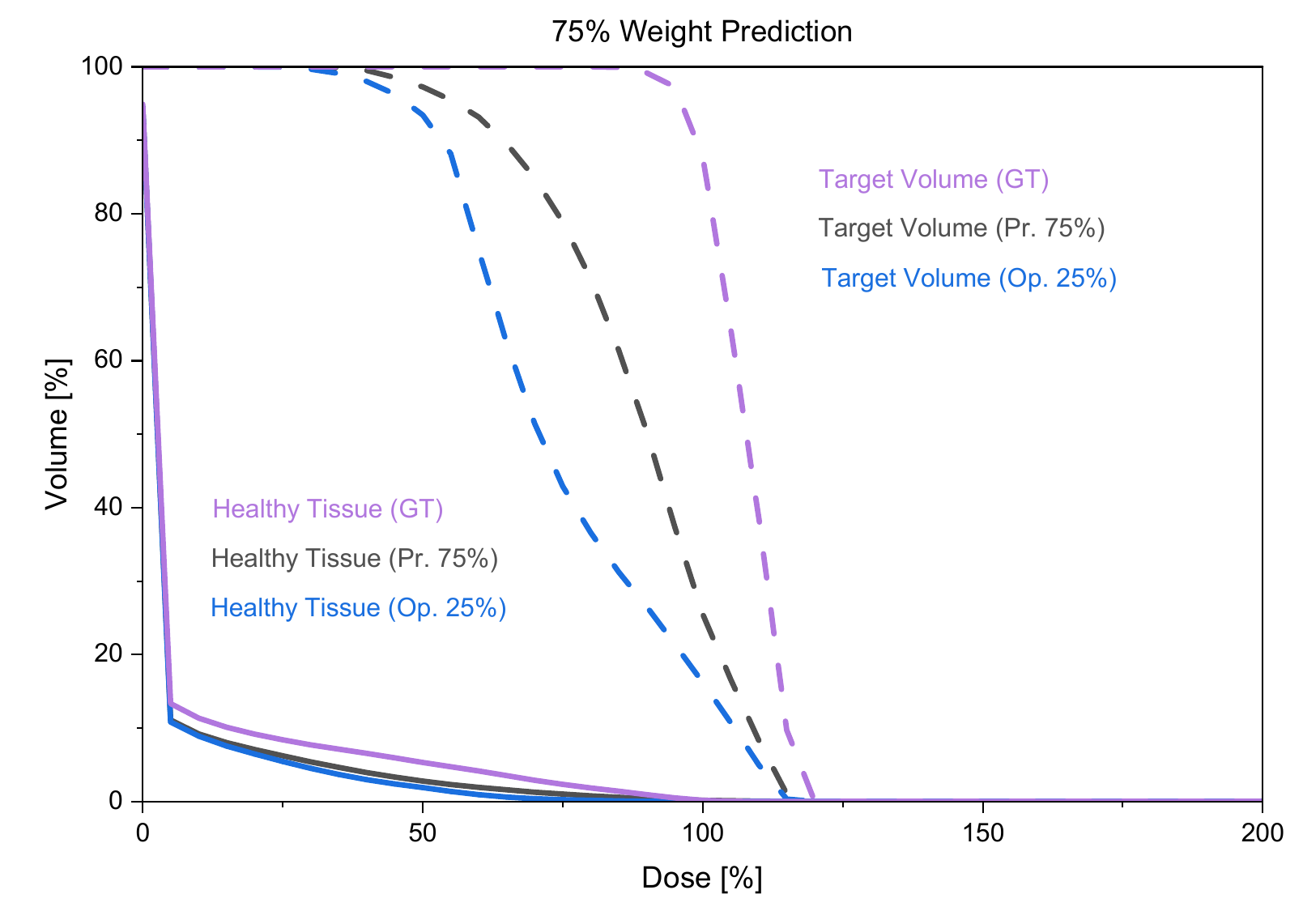}
        \caption{}\label{fig:DVH75}
    \end{subfigure}%
    \hfill 
    \begin{subfigure}{0.5\textwidth}
        \centering
        \includegraphics[width=\textwidth]{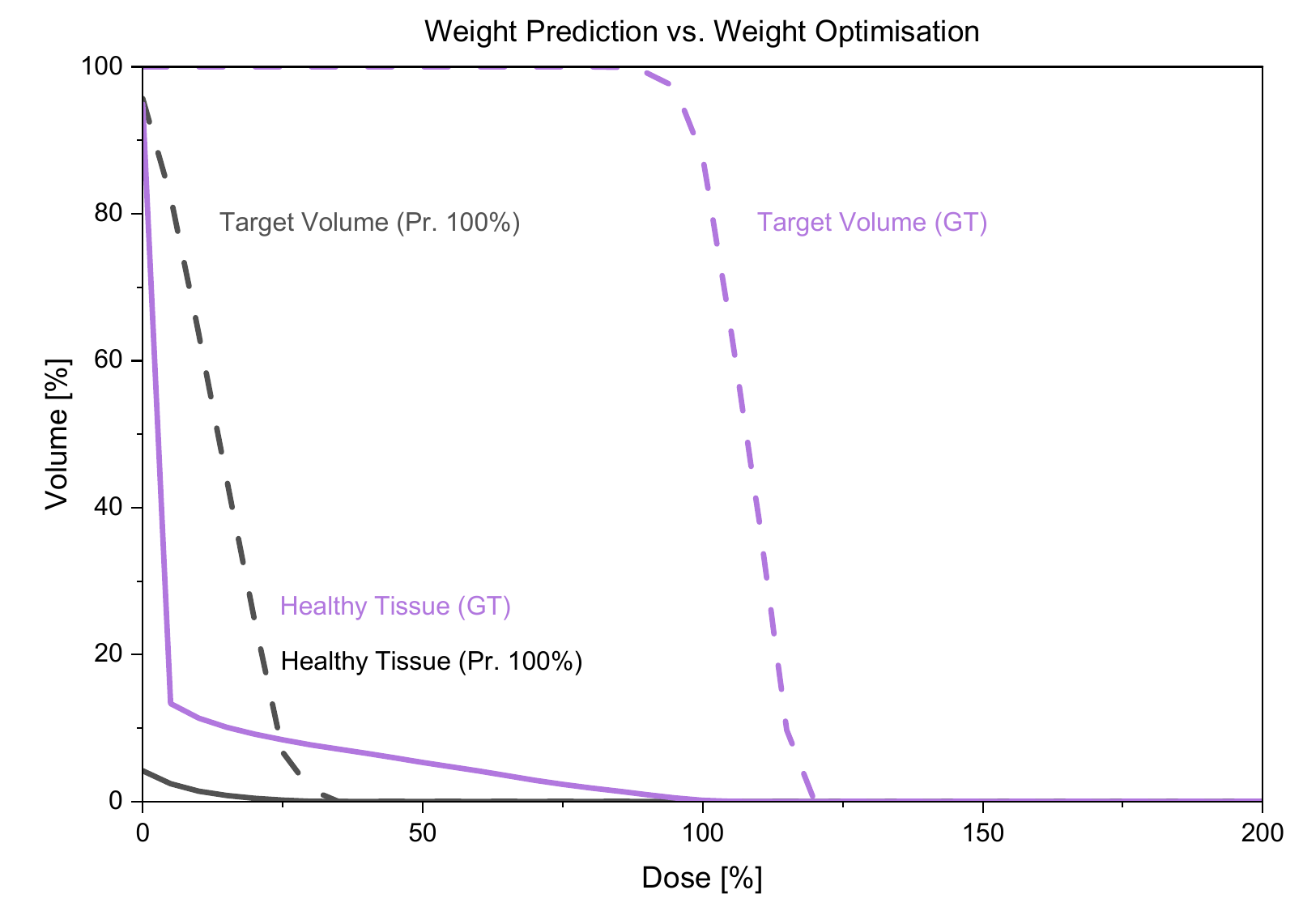}
        \caption{}\label{fig:DVH100}
    \end{subfigure}
    \caption{Dose-volume histograms (DVHs) for the treatment plans shown in Figure~\ref{fig:results}. Figures~\ref{fig:DVH25}-\ref{fig:DVH75}: The lilac curve represents the fully optimised reference plan, the blue curve the accumulated dose at the prediction stage, and the black curve the hybrid treatment plan obtained after incorporating the predicted beam weights. Figure~\ref{fig:DVH100} compares the DVHs of the fully optimised ground truth plan (lilac) and the treatment plan generated using the fully predicted beam weights (black).} \label{fig:DVH}
\end{figure}

Figure~\ref{fig:DVH} presents the corresponding dose volume histograms (DVHs) for the dose distributions shown in Figure~\ref{fig:results}. Figures~\ref{fig:DVH25}-\ref{fig:DVH75} compare the fully optimised GT plan (lilac), the partially delivered GT plan provided as model input (blue), and the hybrid dose distribution obtained by combining the optimised and predicted beam weights (black). The DVH demonstrate that, in all partial prediction scenarios, the completed treatment plans are closer to the fully optimised reference than the partially delivered GT plans. This indicates that the predicted beam weights consistently improve the dosimetric quality of the partially delivered plans, irrespective of the prediction stage. Figure~\ref{fig:DVH100} compares the DVHs of the fully optimised GT plan and the treatment plan generated using the fully predicted beam weights.

\section{Discussion}
In this work, we introduced a conditional hybrid ResNet18–Transformer model for predicting pencil beam weights in 4D treatment planning. It can predict partial beam weights while maintaining the previously optimised dose distribution and could serve as an extension of the 4D treatment planning workflow for treatment plan adaptation. A key aspect of the proposed approach is the use of a physics-based loss function. Instead of minimising the difference between predicted and optimised beam weights, the network is trained through the resulting dose distribution using clinically relevant metrics. This allows the model to prioritise the final treatment objective, namely homogeneous target coverage, while limiting dose to healthy tissue, rather than learning a specific optimised solution. Consequently, the model is not required to replicate the exact optimisation process; instead, it learns to generate beam weights that satisfy the same constraints.

The Transformer architecture enables prediction of either the complete treatment plan or only the remaining beam weights from an intermediate stage of delivery. This capability is particularly relevant for iterative treatment planning, where beam weights may be updated after partial delivery to account for changes in the treatment state. The results demonstrate that the model can improve partially delivered treatment plans by predicting the remaining beam weights while retaining the previously delivered dose contribution.

The prediction accuracy improved as the number of beam weights to be predicted decreased. The best agreement with the fully optimised reference plans was achieved when predicting only the final fraction of the treatment delivery, while predicting the complete treatment plan resulted in larger deviations. This can be seen both in the dose distribution in Figure~\ref{fig:results} and the corresponding DVHs in Figure~\ref{fig:DVH}. This behaviour is likely related to two factors. First, later delivered beams generally contribute less to the overall dose distribution, as the majority of the prescribed dose has already been delivered. Therefore, predicting these remaining beam weights represents a less challenging task. Second, the training distribution is inherently biased towards later treatment stages, as the model always receives queries corresponding to the final part of the delivery sequence. Whereas the earlier sections are queried much less frequently. Consequently, the network predicts the later stages of delivery more frequently than the early stages, which involve spots that contribute more significantly to the treatment.

Future work could investigate strategies to improve prediction performance for earlier treatment stages. One possible approach would be weighted sampling to increase the frequency of challenging early-stage prediction scenarios. Another approach would be to introduce intermediate training tasks, where the network predicts earlier beam weights while later beam weights remain fixed to their optimised values. Although this does not directly correspond to the temporal delivery process, it could allow the network to learn the more challenging initial beam weight predictions before being tasked with predicting the complete delivery sequence.

Several limitations of this study should be considered. First, the evaluation was performed exclusively on simulated phantom patient data. While this approach provides complete control over the 4D treatment-planning workflow and enables reproducible comparison with the optimisation framework, the dataset's anatomical diversity remains limited. 
The current dataset contains similar structures and does not capture the full range of patient-specific variations encountered in clinical practice. Future work should therefore evaluate the proposed model on a larger and more anatomically diverse dataset, including clinical patient data where available. 
Second, the generalisability of the model to different scanners, delivery systems, and machine-specific parameters remains to be investigated. Since delivery times are determined by machine-specific scanner parameters, adaptation to other delivery systems may require additional training data or transfer learning.

Despite these limitations, a major advantage of the proposed approach is the potential substantial reduction in computational cost. While conventional beam weight optimisation requires several hours, the neural network prediction can be performed within milliseconds. This speed-up would enable repeated plan evaluation and refinement, which is impractical when relying solely on conventional optimisation. Therefore, fast beam weight prediction represents a promising direction towards iterative 4D treatment planning workflows, where plans can be continuously updated, simplified, and evaluated while maintaining a homogeneous dose coverage of the target and minimising radiation exposure to healthy tissue.

\section{Conclusion}
This work presented a conditional hybrid ResNet18–Transformer model for predicting pencil beam weights in 4D proton treatment planning. The Transformer architecture supports both complete and partial beam weight prediction, enabling efficient updates to treatment plans at arbitrary stages of delivery. 
Overall, this work demonstrates the feasibility of conditional, physics-informed beam weight prediction for partial treatment replanning. While prediction of complete treatment plans remains an open challenge, the proposed framework provides a promising foundation for future learning-based alternatives to conventional beam weight optimisation.

\newpage

\bibliographystyle{splncs04}
\bibliography{tretmentoptimisation}

\end{document}